\documentclass[a4paper,10pt]{article}
\pdfoutput=1 

\usepackage{jcappub} 
 \usepackage{indentfirst}
\usepackage{appendix}
\usepackage{color}

\title{Exploration of interacting dynamical dark energy model with interaction term including the equation-of-state parameter: alleviation of the $H_{0}$ tension}

\author[a,1]{Rui-Yun Guo\note{Corresponding author.},}
\affiliation[a]{School of Sciences, Xi'an Technological University, Xi'an 710021, China}
\author[b]{Lu Feng,}
\affiliation[b]{College of Physical Science and Technology, Shenyang Normal University, Shenyang 110034, China}
\author[a]{Tian-Ying Yao,}
\author[a]{Xing-Yu Chen}

\emailAdd{guoruiyun110@163.com}
\emailAdd{fengluu@foxmail.com}
\emailAdd{yaotianying0707@163.com}
\emailAdd{1369472231@qq.com}

\abstract{We explore a scenario of interacting dynamical dark energy model with the interaction term $Q$ including the varying equation-of-state parameter $w$. Using the data combination of the cosmic microwave background, the baryon acoustic oscillation, and the type Ia supernovae, to global fit the interacting dynamical dark energy model, we find that adding a factor of the varying $w$ in the function of $Q$ can change correlations between the coupling constant $\beta$ and other parameters, and then has a huge impact on the fitting result of $\beta$. In this model, the fitting value of $H_{0}$ is lower at the $3.54\sigma$ level than the direct measurement value of $H_{0}$. Comparing to the case of interacting dynamical dark energy model with $Q$ excluding $w$, the model with $Q$ including the constant $w$ is more favored by the current mainstream observation. To obtain higher fitting values of $H_{0}$ and narrow the discrepancy of $H_{0}$ between different observations, additional parameters including the effective number of relativistic species, the total neutrino mass, and massive sterile neutrinos are considered in the interacting dynamical dark energy cosmology. We find that the $H_{0}$ tension can be further reduced in these models, but is still at the about $3\sigma$ level.  }


\begin{document}
\maketitle
\flushbottom

\section{Introduction}\label{sec1}

Currently, the direct measurement value of Hubble constant ($H_{0}$) is larger at about $4\sigma\sim6\sigma$ level than the fitting value of $H_{0}$ derived from the Planck observation \cite{Aghanim:2018eyx,Riess:2020fzl}. The discrepancy of $H_{0}$ between different observations has aroused great attention in cosmology and astronomy. For explaining the physical origin behind this problem, some people have made great efforts to reduce the possible systematic errors in the either Planck or SH0ES data. But it seems to not have a remarkable reduction to the $H_{0}$ tension \cite{Spergel:2013rxa,Addison:2015wyg,Planck:2016tof,Efstathiou:2013via,Cardona:2016ems,Zhang:2017aqn,
Follin:2017ljs}. On the other hand, ones dedicate to modify the cosmological standard model (the $\Lambda$ cold dark matter model, i.e., the $\Lambda$CDM model)\footnote{A review about the $H_{0}$ tension, see Ref. \cite{DiValentino:2021izs}}. By extending the $\Lambda$CDM model, they wish to obtain a higher fitting value of $H_{0}$, and then reduce the $H_{0}$ tension between different observations.

In the base $\Lambda$CDM cosmology, dark energy is regarded as vacuum energy with its equation-of-state parameter $w=-1$. In addition to vacuum energy, dark energy may also be a field of dynamical evolution, where $w$ evolves with $z$. When the dynamical dark energy is considered, the results of Refs. \cite{Freese:2004vs,Li:2013dha,Huang:2016fxc,Camarena:2018nbr,Planck:2018vyg,Yang:2021flj,Martinelli:2019krf,
Vagnozzi:2019ezj} have showed that the dynamical dark energy with $w<-1$ at low redshifts prefers a high value of $H_{0}$. Besides, when the interaction between dark energy and dark matter is considered, the $H_{0}$ tension also can be relieved to some extent \cite{DiValentino:2017iww,Yang:2017ccc,Feng:2017usu,Yang:2018euj,An:2018vzw,Wang:2016lxa,Yang:2021oxc,
Yang:2021hxg,Yang:2018xlt,Yang:2017zjs,Li:2019ajo,DiValentino:2019ffd,DiValentino:2019jae,DiValentino:2020kpf,
Zhang:2021yof,Bonilla:2021dql,Johnson:2021wou}. However, neither the dynamical dark energy nor the interacting dark energy can relieve the $H_{0}$ tension at less than $2\sigma$ level. Usually, to obtain a higher fitting value of $H_{0}$, some extra parameters are introduced in cosmological models. These parameters include the effective number of relativistic species ($N_{\rm eff}$), the total neutrino mass ($\sum m_{\nu}$), massive sterile neutrinos ($N_{\rm eff}$ and $m_{\nu, \rm sterile}^{\rm eff}$), and other extra parameters, which have potential to provide higher fitting values of $H_{0}$ \cite{Guo:2018gyo,Battye:2013xqa,Bernal:2016gxb,Guo:2017hea,Guo:2017qjt,Yang:2020ope,Yang:2020tax,DiValentino:2017rcr,SDSS:2014iwm,Guo:2018ans,Zhao:2017urm,Feng:2019jqa,Poulin:2018cxd,
Sakstein:2019fmf}. But so far, the problem of the $H_{0}$ tension has still not been completely explained.

Although the above extensions to the $\Lambda$CDM model can not provide a fitting value of $H_{0}$ close to the direct measurement value of $H_{0}$, these researches show that considering extra parameters in cosmological models can indeed increase the fitting value of $H_{0}$. Thus, in this paper we revisit these situations. We first explore the impact of the interacting dark energy on the fitting result of $H_{0}$. In cosmology, the cosmological model that considers the interaction between dark energy and dark matter is usually called the interacting dark energy model, abbreviated as ``IDE" model. In IDE models, the energy conservation equations for dark energy and cold dark matter satisfy
\begin{equation}\label{1.1}
  \dot{\rho}_{\rm de}=-3H(1+w)\rho_{\rm de}+Q_{\rm de},
\end{equation}
\begin{equation}\label{1.2}
  \dot{\rho}_{\rm c}=-3H\rho_{\rm c}+Q_{\rm c},
\end{equation}
where the overdot always denotes the derivative with respect to the cosmic time $t$, $\rho_{\rm c}$ and $\rho_{\rm de}$ denote the density of cold dark matter and dark energy, $H$ is Hubble parameter, and the energy transfer rate $Q=Q_{\rm de}=-Q_{\rm c}$. $Q>0$ denotes that cold dark matter decays into dark energy, $Q<0$ denotes that dark energy decays into cold dark matter, and $Q=0$ indicates no interaction between dark energy and dark matter.

According to equations. (\ref{1.1}) and (\ref{1.2}), the forms of $w$ and $Q$ should be given for a complete IDE model. The fitting value of $H_{0}$ is influenced by the forms of $w$ and $Q$ in IDE models. $w$ is anti-correlated with $H_{0}$, and the coupling constant $\beta$ ($Q=\beta H\rho_{\rm c}$) is positively correlated with $H_{0}$ \cite{Guo:2018ans}. However, there is currently no theory that can determine the specific form of $Q$ from the first principles, people can only construct a reasonable interaction model phenomenologically. In past research, people usually take the form of $Q\propto \beta \rho_{i}$ ($i$ denotes ``c" or ``de") \cite{Guo:2018gyo,Guo:2017hea,Li:2015vla,Li:2014cee,Yang:2021oxc,Anchordoqui:2021gji}. In the case of $Q\propto \beta \rho_{\rm c}$, the coupling constant $\beta$ and $H_{0}$ has a strong correlation. Also, the fitting result of $Q\propto \beta \rho_{\rm c}$ is more tighter than the case of $Q\propto \beta \rho_{\rm de}$ \cite{Guo:2018gyo,Guo:2017hea,Li:2015vla,Li:2014cee}. Thus, in this paper we only consider the case of $Q\propto\beta \rho_{\rm c}$.

It is generally believed that the dark energy that can interacts with dark matter should be a dynamic field or fluid. The state equation of dark energy $w$ should be a constant or a form that evolves with time.
In Ref. \cite{Pan:2019gop}, a factor of $(1+w)$ is added in the function of $Q$, i.e., $Q=\beta H (1+w) \rho_{\rm c}$. Their results showed that the fitting value of $H_{0}=71.70^{+1.50}_{-1.70}$ km/s/Mpc is higher compared to the $\Lambda$CDM-based Planck's fitting result. In the model with $Q=\beta H (1+w) \rho_{\rm c}$, the fitting results of $H_{0}$ \cite{Pan:2019gop} are close to the local measurements of $H_{0}$, thus alleviating the $H_{0}$ tension. Inspired by \cite{Pan:2019gop}, the function of $Q=\beta H (1+w) \rho_{\rm c}$ can be split into two parts of $Q=\beta H \rho_{\rm c}$ and $Q=\beta H w \rho_{\rm c}$ (In fact, the former is a special case of the latter). When $w=-1$ and $w=constant$, the IDE model with $Q=\beta H \rho_{\rm c}$ can be called ``IDE1" and ``IDE1+$w$". The IDE1 model denotes that the vacuum energy interacts with dark matter. In this case, though $w=-1$, the vacuum energy density becomes a dynamical quantity because of its perturbation as the response to the metric fluctuations. The IDE1+$w$ model denotes that the dynamical dark energy with a constant $w$ interacts with dark matter. The two cases of $Q=\beta H \rho_{\rm c}$ with $w=-1$ and $w=constant$ have been widely discussed to resolve the problem of the $H_{0}$ tension~\cite{Valiviita:2008iv,DiValentino:2019ffd,Yang:2021oxc,DiValentino:2019jae,Benetti:2019lxu,Yang:2019uog}.

In this paper, the IDE model with $Q=\beta H w \rho_{\rm c}$ and $w=constant$ will be abbreviated as ``IDE2". The model also denotes that the dynamical dark energy with a constant $w$ interacts with dark matter. It seems that it has similar cosmological (background and perturbation) evolution to that in the IDE1+$w$ model. But in this case, the function of $Q$ is phenomenologically constructed to be proportional to $\rho_{\rm c}$ (the density of dark matter) and $w$ (characterizing the nature of dark energy). The fitting results of cosmological parameters in the IDE model may be influenced by different forms of $Q$, thus we revisit the constraints on the IDE2 model with the form of $Q=\beta H w \rho_{\rm c}$. Comparing to the above IDE models with $Q=\beta H \rho_{\rm c}$ (i.e., the IDE1 model and the IDE1+$w$ model), we wish the IDE2 model with $Q=\beta H w \rho_{\rm c}$ to be more consistent with the direct measurement value of $H_{0}$. Actually, the form of $Q=\beta H w \rho_{i}$ has been proposed \cite{Zhang:2004gc,Wang:2013qy,Bento:2002yx}, but they did not get enough consideration in past research.

In IDE cosmology, dark energy interacts with dark matter each other. Under these circumstances, the curvature perturbation is in the rapid and unlimited growth in the early universe. This is so-called ``large-scale instability problem" \cite{Majerotto:2009zz,Clemson:2011an,He:2008si,Valiviita:2008iv,Hu:2008zd,Fang:2008sn}. For the scenario of $Q\propto \rho_{\rm c}$, the large-scale instability will appear under the value of $w>-1$. In order to resolve the large-scale instability problem, a new framework for calculating the cosmological perturbation of interacting dark energy using the parameterized post-Friedmann (PPF) method has been developed. This method has been confirmed to be able to eliminate the large-scale instability of interacting dark energy very well \cite{Guo:2017hea,Feng:2017usu,Li:2015vla,Li:2014eha,Li:2014cee}.  For more details about the PPF method, we refer the reader to refs. \cite{Li:2014eha,Li:2014cee}.

To obtain a high $H_{0}$ and narrow the discrepancy of $H_{0}$ between different observations, we will further introduce some other parameters including $\sum m_{\nu}$, $N_{\rm eff}$, and $m_{\nu, \rm sterile}^{\rm eff}$. These parameters can influence the fitting value of $H_{0}$. Using the current mainstream observations to global fit the IDE1, IDE1+$w$, IDE2, and their extensive models, our main aims are to investigate these questions in the following: (i) After adding a factor of $w$ in the function of $Q=\beta H \rho_{\rm c}$, what fitting results of cosmological parameters will be obtained? (ii) Compared with the IDE1+$w$ model, whether the IDE2 model will favor a higher fitting value of $H_{0}$ ? (iii) Whether the $H_{0}$ tension can be relieved by considering extra parameters in the IDE model? The organization of this paper is to describe the data and method used in this paper in Sec. \ref{sec:2}, to analysis and discuss the fitting results of cosmological parameters obtained in this paper in Sec. \ref{sec:3}, and to make a conclusion finally in Sec. \ref{sec:4}.

\section{Data and method}\label{sec:2}

We employ the combination of the cosmic microwave background (CMB) data, the baryon acoustic oscillation (BAO) data, and the type Ia supernovae (SNe) data, i.e., CMB+BAO+SNe. They are the current mainstream observational data. In the following, these data will be described in detail.
\begin{itemize}
  \item \emph{\textbf{CMB}}: The CMB data include the combined likelihood of the Planck temperature and polarization power spectra at $\ell \geq30$, the low$-\ell$ temperature Commander likelihood and the SimAll EE likelihood, together with the lensing power spectrum data, from the 2018 Planck data release \cite{Aghanim:2018eyx}.
      Compared to the previous CMB data, these data are highly efficient and robust.
  \item \emph{\textbf{BAO}}: We employ the 6dF Galaxy Survey (6dFGS) and Main Galaxy Sample
of Data Release 7 of Sloan Digital Sky Survey (SDSS-MGS) measurements of the acoustic-scale distance
ratio $D_{\rm V}/r_{\rm drag}$~\cite{Beutler:2011hx,Ross:2014qpa}, together with the final Data Release 12 (DR12) BAO results~\cite{Alam:2016hwk} in three redshift slices with effective redshifts $z_{\rm eff}=0.38$, $0.51$, and $0.61$. Here, $r_{\rm drag}$ is the comoving sound horizon at the end of the baryon drag epoch, and $D_{V}$ is a combination of the comoving angular diameter distance $D_{M}(z)$ and Hubble parameter $H(z)$.
  \item \emph{\textbf{SNe}}: We use the ``Pantheon'' sample~\cite{Scolnic:2017caz} for supernovae, which contains 1048 supernovae samples within the redshift range of $0.01<z<2.3$. These data are constructed from 276 supernovae from the Pan-STARRS1 Medium Deep Survey at $0.03 < z < 0.65$ plus several samples of low redshift and HST. The Pantheon data can provide tighter constraints on cosmological parameters than the ``Joint Light-curve Analysis" (JLA) analysis~\cite{Aghanim:2018eyx}.

\end{itemize}

The $\chi^2$ statistic is adopted to fit the cosmological models to observational data. The total $\chi^2$ of the CMB+BAO+SNe data can be written as
\begin{equation}
\chi^2=\chi^{2}_{\rm{CMB}}+\chi^{2}_{\rm{BAO}}+\chi^{2}_{\rm{SNe}}.
\end{equation}
For every observation, the $\chi^2$ function is defined by
\begin{equation}
\chi^2_{\xi}=\frac{(\xi_{\rm {th}}-\xi_{\rm {obs}})^{2}}{\sigma^{2}_{\xi}},
\end{equation}
where $\xi_{\rm{th}}$, $\xi_{\rm{obs}}$, and $\sigma_{\xi}$ denote the theoretically predicted value, the experimentally measured value, and the standard deviation, respectively. For different cosmological models with different numbers of parameters, a model with more parameters has a more preference for a lower value of $\chi^2$. Given this fact, we also apply the rather popular Akaike information criterion (AIC) to do the fair model comparison.

We have
\begin{equation}
{\rm AIC}=-2\ln{\mathcal{L}_{\rm{max}}}+2k,
\end{equation}
where $\mathcal{L}_{\rm max}$ and $k$ are the maximum likelihood and the number of parameters. For Gaussian errors, $\chi^{2}_{\rm{min}}=-2\ln{\mathcal{L}_{\rm{max}}}$. In practice, the relative values between different models are more applicable and valuable, i.e., we have $\Delta {\rm AIC}=\Delta\chi^{2}_{\rm{min}}+2\Delta k$. A model with a lower AIC value is more favored by data. Roughly speaking, compared to a reference model, the models with $0<\Delta {\rm AIC}<2$ have substantial support, the models with $4<\Delta {\rm AIC}<7$ have considerably less support, and the models with $\Delta {\rm AIC}>10$ have essentially no support~\cite{H.Akaike:1974}.

In a spatially flat universe, we assume the $\Lambda$CDM model as a reference model. Following the Planck 2018 release, we put the same flat priors on the baryon density $\omega_{\rm b}\equiv \Omega_{\rm b}h^{2}$, cold dark matter density $\omega_{\rm c}\equiv \Omega_{\rm c}h^{2}$, an approximation to the observed angular size of the sound horizon at recombination $\theta_{\rm MC}$, the reionization optical depth $\tau$, the initial super-horizon amplitude of curvature perturbations $A_{\rm s}$ at $k=0.05$ Mpc$^{-1}$, and the primordial spectral index $n_{\rm s}$. For the IDE1 model, the additional free parameter is $\beta$ with the prior of $[-0.3, 0.3]$. The IDE1+$w$ model and the IDE2 model have the same number of parameters, and the additional free parameters are $\beta$ with the prior of $[-0.3, 0.3]$ and $w$ with the prior of $[-3, 1]$. For the extra parameter $\sum m_{\nu}$, we assume a normal neutrino mass hierarchy with the minimal mass $\sum m_{\nu}=0.06$ eV. When only the parameter $N_{\rm eff}$ is considered, its prior is set as $[0,6]$. But when a massive sterile neutrino is considered, the priors are $[3.046, 7]$ for $N_{\rm eff}$ and $[0, 10]$ eV for $m_{\nu, \rm sterile}^{\rm eff}$.

Our main results are based upon the CMB+BAO+SNe data computed with the August 2017 version of the {\tt camb} Boltzmann code~\cite{Lewis:1999bs}, and parameter constraints are based on the July 2018 version of {\tt CosmoMC}~\cite{Lewis:2013hha}. By modifying and running this code package, we can obtain the posterior distributions of parameters, the best-fit points with $\chi^{2}_{\rm min}$, and $1\sigma$ and $2\sigma$ boundaries, etc. For more details of the calculation methods, we refer the reader to Refs.~\cite{Lewis:1999bs,Lewis:2013hha}.

\section{Results and discussion}\label{sec:3}

\begin{table*}[!htp]
\centering
\renewcommand{\arraystretch}{1.5}
\scalebox{0.9}[0.9]{%
\begin{tabular}{|c| c c c |}
\hline
 Model&IDE1 &IDE1+$w$ &IDE2 \\
\hline

$\beta$
                 &$0 .0003\pm0.0012$
                 &$-0.0011^{+0.0014}_{-0.0015}$
                 &$0.0010^{+0.0015}_{-0.0013}$
                     \\

$w$
                                              &$-1$
                                              &$-1.043^{+0.036}_{-0.037}$
                                              &$-1.041^{+0.037}_{-0.036}$
                                              \\
\hline

$H_0$ [km/s/Mpc]($\delta H_{0}$)
                                              &$67 .86\pm0.63$($3.97\sigma$)
                                              &$68 .26^{+0.83}_{-0.82}$($3.51\sigma$)
                                              &$68 .23^{+0.82}_{-0.83}$($3.54\sigma$)
                                              \\

$S_{8}$     &$0 .824\pm0.010$
            &$0 .827\pm0.011$
            &$0 .827\pm0.011$
\\

\hline

$\chi^{2}_{\rm min}$
                                              &$3824.268$
                                              &$3823.888$
                                              &$3822.360$
\\

$\Delta \rm AIC$  &$1.346$
                                              &$2.966$
                                              &$1.438$
                                              \\
\hline
\end{tabular}}
\caption{ Fitting results of cosmological parameters in the IDE1, IDE1+$w$, and IDE2 models using the CMB+BAO+SNe data combination. }
\label{tab:l}
\end{table*}

\begin{figure*}[ht!]
\begin{center}
\includegraphics[width=15.0cm]{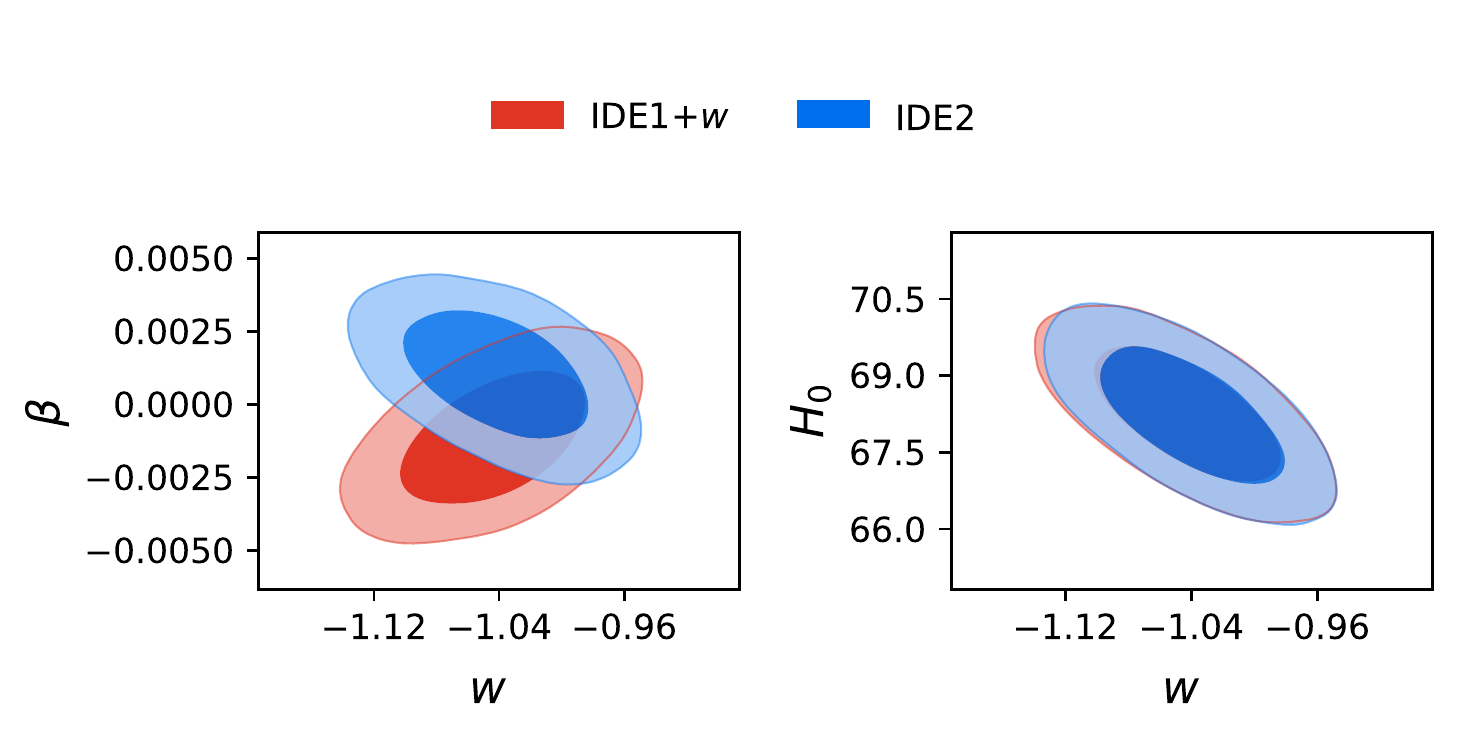}
\end{center}
\caption{ Two-dimensional marginalized contours ($1\sigma$ and $2\sigma$) in the $w$--$\beta$ plane and the $w$--$H_{0}$ plane for the IDE1+$w$ and IDE2 models by using the CMB+BAO+SNe data combination.}
\label{fig1}
\end{figure*}

\begin{figure*}[ht!]
\begin{center}
\includegraphics[width=15.0cm]{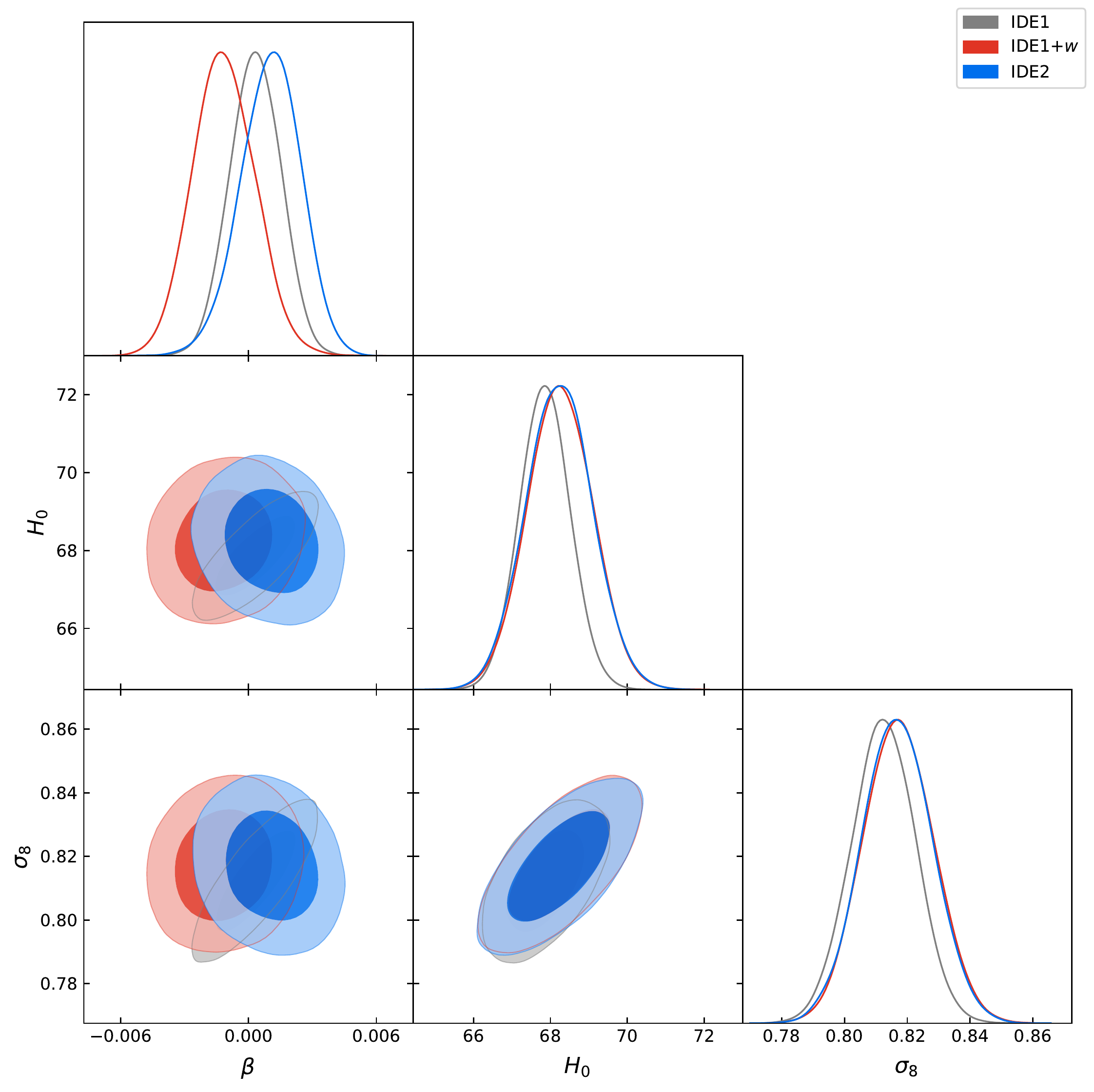}
\end{center}
\caption{The one-dimensional posterior distributions and two-dimensional marginalized contours ($1\sigma$ and $2\sigma$) for the parameters $\beta$, $H_{0}$, and $\sigma_{8}$ in the IDE1, IDE1+$w$, and IDE2 models by using the CMB+BAO+SNe data combination.}
\label{fig2}
\end{figure*}

Firstly, we give the constraint results of the IDE1, IDE1+$w$, and IDE2 models in the Tab.\ref{tab:l}. We obtain $\beta=0 .0003\pm0.0012$ in the IDE1 model, $\beta=-0.0011^{+0.0014}_{-0.0015}$ in the IDE1+$w$ model, and $\beta=0.0010^{+0.0015}_{-0.0013}$ in the IDE2 model. $\beta=0$ is favored at the $1\sigma$ level by the CMB+BAO+SNe data combination, indicating that there is no evidence of a nonzero interaction. When the dynamical dark energy with a constant $w$ is considered in the IDE model, the fitting central value of $\beta$ has a relatively large deviation. In the IDE1+$w$ model, the fitting central value of $\beta$ becomes smaller than that in the IDE1 model with $w=-1$. But in the IDE2 model, it becomes larger. This may be due to having different correlations between $w$ and $\beta$ in the IDE1+$w$ model and the IDE2 model, as shown in Fig.\ref{fig1}.

In the IDE1+$w$ model, $Q=\beta H \rho_{\rm c}$ only includes a free parameter $\beta$. Combining $Q=\beta H \rho_{\rm c}$ with the equation. (\ref{1.1}), we can obtain the effective equation-of-state parameter $w_{\rm eff}=w-\frac{1}{3}\beta \frac{\rho_{\rm c}}{\rho_{\rm de}}$. To keep $w_{\rm eff}$ close to a constant, a smaller $w$ leads to a smaller $\beta$. In the IDE2 model, $Q=\beta Hw \rho_{\rm c}$, we obtain $w_{\rm eff}=w-\frac{1}{3}\beta w \frac{\rho_{\rm c}}{\rho_{\rm de}}$. For the fixed $w_{\rm eff}$, the changes of $w$ and $\beta$ should be consistent in theory. But in this case, $Q=\beta Hw \rho_{\rm c}$ includes two free parameters ($w$ and $\beta$). The fact that the current observational data favor $Q$ close to a constant leads to a strong anti-correlation between $w$ and $\beta$. Thus $w$ is positively correlated with $\beta$ in the IDE1+$w$ model, but is anti-correlated with $\beta$ in the IDE2 model, qualitatively. This indicates that adding a factor of $w$ in the function of $Q$ can change the correlation between $w$ and $\beta$, thus affecting the fitting result of $\beta$. We obtain $w=-1.043^{+0.036}_{-0.037}$ in the IDE1+$w$ model and $w=-1.041^{+0.037}_{-0.036}$ in the IDE2 model, meaning that $w<-1$ (the phantom dark energy) is favored within the $1\sigma$ range.

For the constraints on the parameter $H_{0}$, we obtain $H_{0}=67 .86\pm0.63$ km/s/Mpc in the IDE1 model, $H_{0}=68 .26^{+0.83}_{-0.82}$ km/s/Mpc in the IDE1+$w$ model, and $H_{0}=68 .23^{+0.82}_{-0.83}$ km/s/Mpc in the IDE2 model. Comparing to the fitting result of $H_{0}=67 .72\pm0.41$ km/s/Mpc in the base $\Lambda$CDM model, a larger fitting value of $H_{0}$ is favored in these IDE models, reducing the $H_{0}$ tension from $4.27\sigma$ to $3.97\sigma$ in the IDE1 model, from $4.27\sigma$ to $3.51\sigma$ in the IDE1+$w$ model, and from $4.27\sigma$ to $3.54\sigma$ in the IDE2 model\footnote{In order to make a comparison with the results of \cite{Aghanim:2018eyx}, we adopt the direct measurement value of $H_{0}=74.03\pm1.42$km/s/Mpc \cite{Riess:2019cxk} to calculate the tension with the fitting value of $H_{0}$ derived from the CMB+BAO+SNe data combination.}. Considering the dynamical dark energy with the constant $w$ in the IDE model seems to relieve the $H_{0}$ tension better. But specifically, the dynamical dark energy model with $Q=\beta Hw \rho_{\rm c}$ is not better than that with $Q=\beta H \rho_{\rm c}$ for the fitting value of $H_{0}$.

For the parameters $\beta$, $H_{0}$, and $\sigma_{8}$ in the IDE1, IDE1+$w$, and IDE2 models, the one-dimensional posterior distributions and two-dimensional marginalized contours ($1\sigma$ and $2\sigma$) are given in Fig.\ref{fig2}. We see the correlations between $\beta$ and other parameters will be changed once adding a factor of $w$ in the function of $Q=\beta H \rho_{\rm c}$. The correlation between $w$ and $H_{0}$ is anti-correlated in Fig.\ref{fig1}, which indicates considering the dynamical dark energy with the constant $w$ inevitably increases the fitting value of $H_{0}$, and further relieves the $H_{0}$ tension.

However, compared with the base $\Lambda$CDM model with $\chi^{2}=3824.922$, the IDE1 model has $\Delta\chi^{2}=-0.654$ and $\Delta \rm AIC=1.346$, the IDE1+$w$ model has $\Delta\chi^{2}=-1.034$ and $\Delta \rm AIC=2.966$, and the IDE2 model has $\Delta\chi^{2}=-2.562$ and $\Delta \rm AIC=1.438$, indicating that the three models are favored by the CMB+BAO+SNe data combination. Among them the IDE1 model and the IDE2 model are more consistent with the current observational data from a statistical point of view, but the fitting value of $H_{0}$ in the IDE1 model is smaller than that in the IDE2 model. Thus, for the three IDE models, the IDE2 model is most economical and effective to relieve the $H_{0}$ tension.

\begin{table*}[!htp]
\centering
\renewcommand{\arraystretch}{1.5}
\scalebox{0.9}[0.9]{%
\begin{tabular}{|c| c c c c|}
\hline
 Model&IDE2+$N_{\rm eff}$ &IDE2+$\sum m_{\nu}$ &IDE2+$N_{\rm eff}$+$\sum m_{\nu}$&IDE2+$N_{\rm eff}$+$m_{\nu,\rm sterile}^{\rm eff}$  \\
\hline

$\beta$
                 &$0.0009^{+0.0015}_{-0.0013}$
                 &$0.0005^{+0.0009}_{-0.0008}$
                 &$0 .0005^{+0.0009}_{-0.0008}$
                 &$0.0008^{+0.0013}_{-0.0012}$
                     \\

$w$
                                              &$-1.047^{+0.037}_{-0.038}$
                                              &$-1.047\pm0.036$
                                              &$-1.053\pm0.038$
                                              &$-1.050\pm0.037$
                                              \\
$N_{\rm eff}$                           &$2.950\pm0.180$
                                              &$3.046$
                                              &$2.960\pm0.180$
                                              &$<3.308$
                                              \\

$\sum m_{\nu}$  [eV]                    &$0.060$
&$<0.184$
                                              &$<0.180$
                                              &$0.060$
                                              \\

$m_{\nu, \rm sterile}^{\rm eff}$   [eV]                   &$0$
                                              &$0$
                                              &$0$
                                              &$<0.589$
                                              \\
\hline

$S_{8}$                  &$0 .825\pm0.011$
                                          &$0 .824\pm0.011$
                                          &$0 .822\pm0.011$
                                          &$0 .812^{+0.021}_{-0.014}$
\\

$H_0$ [km/s/Mpc]($\delta H_{0}$)
                                              &$67.80\pm1.20$($3.34\sigma$)
                                              &$68.26^{+0.82}_{-0.83}$($3.51\sigma$)
                                              &$67.80\pm1.20$($3.34\sigma$)
                                              &$68.64^{+0.87}_{-0.97}$($3.13\sigma$)
                                              \\

\hline

$\chi^{2}_{\rm min}$       &$3821.834$
                                              &$3822.132$
                                              &$3821.382$
                                              &$3821.582$
\\

$\Delta \rm AIC$  &$2.912$
                                              &$3.210$
                                              &$4.460$
                                              &$4.660$
\\
\hline
\end{tabular}}
\caption{Fitting results of cosmological parameters in a range of extensions to the IDE2 model by using the CMB+BAO+SNe data combination.}
\label{tab:2}
\end{table*}

\begin{figure*}[ht!]
\begin{center}
\includegraphics[width=15.0cm]{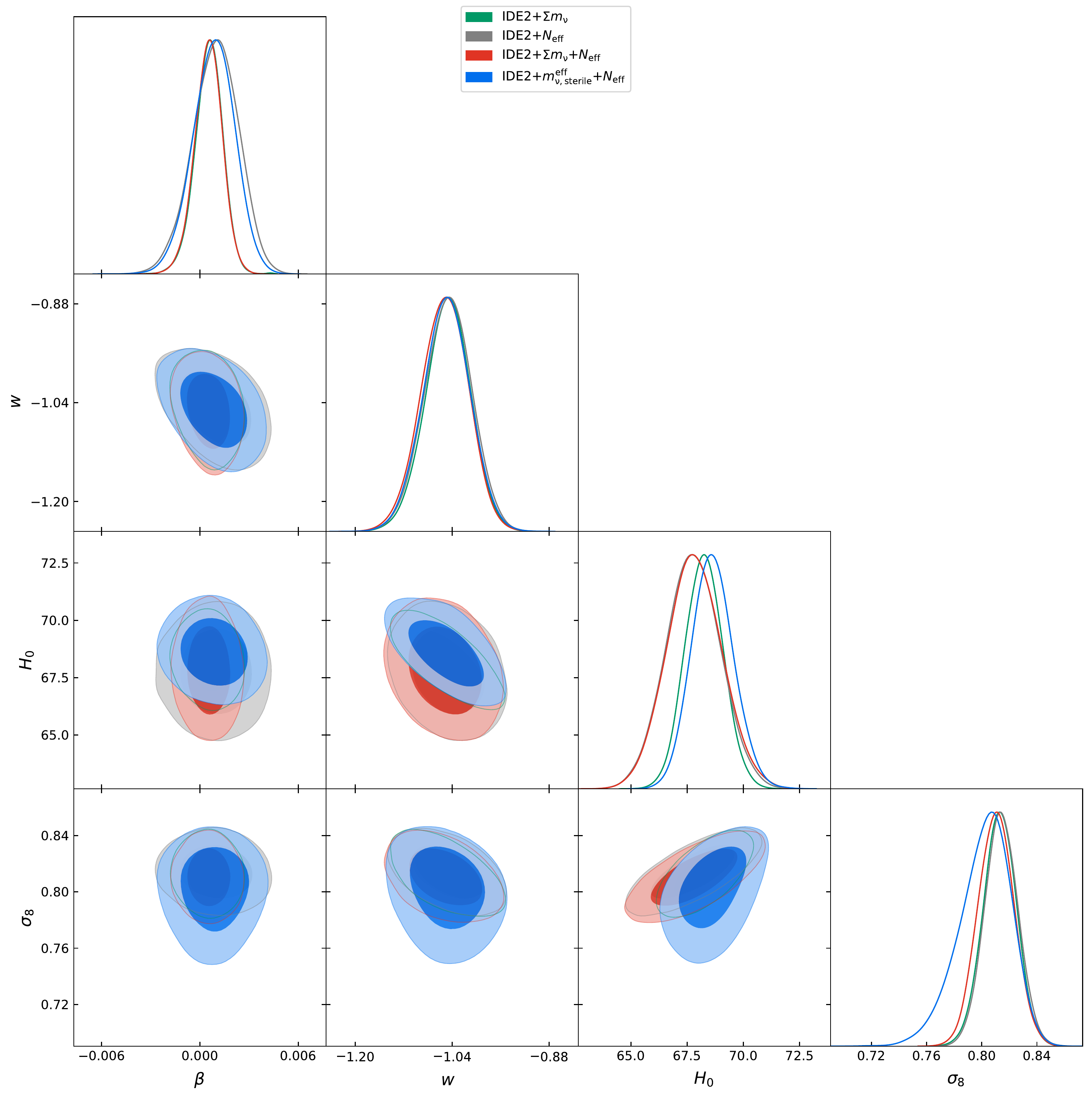}
\end{center}
\caption{The one-dimensional posterior distributions and two-dimensional marginalized contours ($1\sigma$ and $2\sigma$) for the parameters $\beta$, $w$, $H_{0}$, and $\sigma_{8}$ in these extensions to the IDE2 model by using the CMB+BAO+SNe data combination.}
\label{fig3}
\end{figure*}

\begin{figure*}[ht!]
\begin{center}
\includegraphics[width=5.0cm]{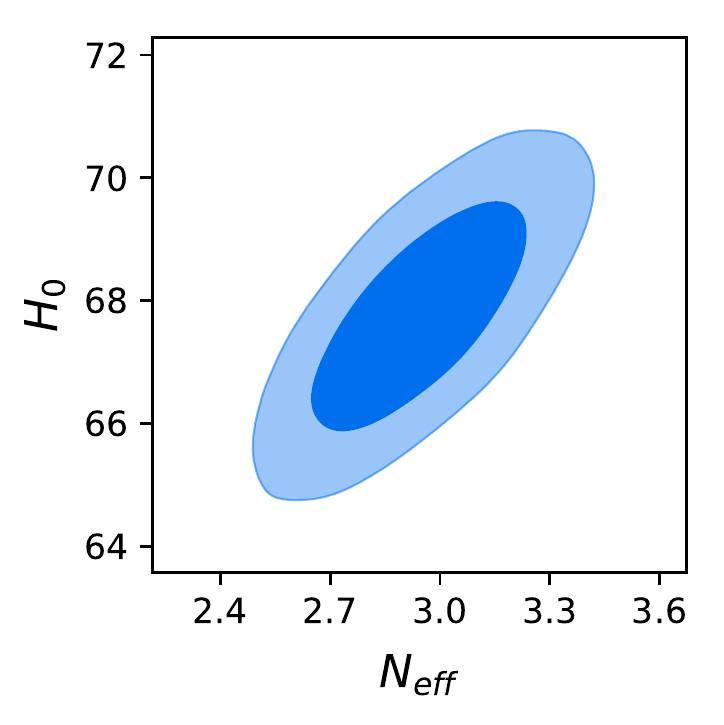}
\includegraphics[width=5.0cm]{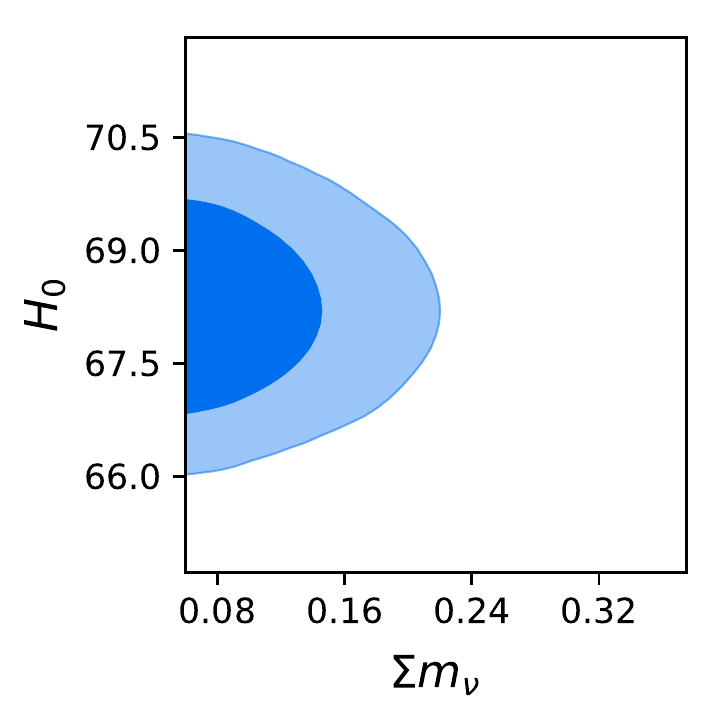}
\includegraphics[width=5.0cm]{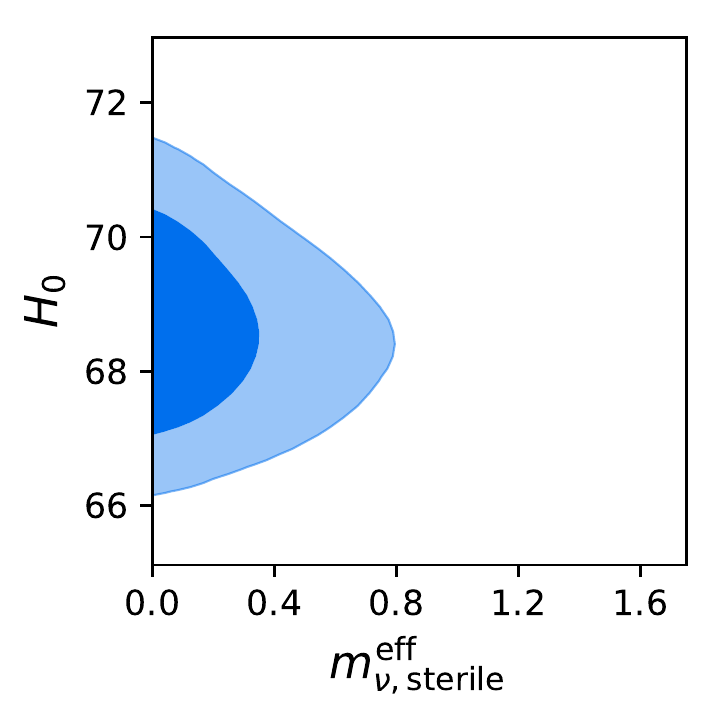}
\end{center}
\caption{Two-dimensional marginalized contours ($1\sigma$ and $2\sigma$) in the $N_{\rm eff}$--$H_{0}$ plane for the IDE2+$N_{\rm eff}$ model, in the $\sum m_{\nu}$--$H_{0}$ plane for the IDE2+$\sum m_{\nu}$ model, and in the $m_{\nu, \rm sterile}^{\rm eff}$--$H_{0}$ plane for the IDE2+$N_{\rm eff}$+$m_{\nu, \rm sterile}^{\rm eff}$ model, by using the CMB+BAO+SNe data combination.}
\label{fig4}
\end{figure*}

Obviously, only considering the case of the IDE model with the constant $w$ is not enough to relieve the $H_{0}$ tension to be at less than $2\sigma$ level. To increase the fitting value of $H_{0}$, we further consider the parameters $N_{\rm eff}$, $\sum m_{\nu}$, and $m_{\nu, \rm sterile}^{\rm eff}$ in the IDE2 model. Adding these special parameters in cosmological models can help to increase the fitting value of $H_{0}$ and reduce the $H_{0}$ tension. We discuss the fitting results of the following four cases: IDE2+$N_{\rm eff}$, IDE2+$\sum m_{\nu}$, IDE2+$N_{\rm eff}$+$\sum m_{\nu}$, and IDE2+$N_{\rm eff}$+$m_{\nu, \rm sterile}^{\rm eff}$. In Tab.\ref{tab:2}, we give the $1\sigma$ fitting results for these extensive models. For the parameters with poor constraints, we employ the $2\sigma$ upper bound. We find that no interaction ($\beta=0$) and the phantom dark energy ($w<-1$) are favored by the current observational data for these extensions to the IDE2 model. Among these models, the parameter $\beta$ in the cases with varying $\sum m_{\nu}$ is constrained better than that with $\sum m_{\nu}=0.06$ eV.

In Fig.\ref{fig3}, we give the one-dimensional posterior distributions and two-dimensional marginalized contours ($1\sigma$ and $2\sigma$) for the parameters $\beta$, $w$, $H_{0}$, and $\sigma_{8}$ in these extensions to the IDE2 model. For the fitting result of $H_{0}$, we obtain $H_{0}=67.80\pm1.20$ km/s/Mpc in the IDE2+$N_{\rm eff}$ model, $H_{0}=68.26^{+0.82}_{-0.83}$ km/s/Mpc in the IDE2+$\sum m_{\nu}$ model, $H_{0}=67.80\pm 1.20$ km/s/Mpc in the IDE2+$N_{\rm eff}$+$\sum m_{\nu}$ model, and $H_{0}=68.64^{+0.87}_{-0.97}$ km/s/Mpc in the IDE2+$N_{\rm eff}$+$m_{\nu,\rm sterile}^{\rm eff}$ model. Correspondingly, the $H_{0}$ tension is relieved to be at the $3.34\sigma$, $3.51\sigma$, $3.34\sigma$, and $3.13\sigma$ level. All extensions to the IDE2 model can help further relieve the $H_{0}$ tension. Among them, the IDE2+$N_{\rm eff}$+$m_{\nu,\rm sterile}^{\rm eff}$ model is the most effective, and the IDE2+$\sum m_{\nu}$ model is worst, to reduce the discrepancy of the $H_{0}$ observations. In the IDE2+$N_{\rm eff}$+$m_{\nu,\rm sterile}^{\rm eff}$ model, we set $\Delta N_{\rm eff}=N_{\rm eff}-3.046>0$. The varying $N_{\rm eff}$ can shift the acoustic peaks in the damping tail of the CMB temperature power spectra, also modify the equal time of matter and radiation. Due to the strong degeneracy between $N_{\rm eff}$ and $H_{0}$~\cite{Aghanim:2018eyx,Guo:2018ans} (also as shown in Fig.\ref{fig4}), it will be possible to obtain a larger fitting value of $H_{0}$. Other three extensive models all favor $\Delta N_{\rm eff}=0$ at the $1\sigma$ level. For the cases of the IDE model with varying $\sum m_{\nu}$, the fitting values of $\sum m_{\nu}$ are larger than that in the $\Lambda$CDM model.

From a statistical point of view, the IDE2+$N_{\rm eff}$+$m_{\nu,\rm sterile}^{\rm eff}$ model with $\Delta \chi^{2}=-3.340$ and $\Delta \rm AIC=4.660$, and the IDE2+$N_{\rm eff}$+$\sum m_{\nu}$ model with $\Delta \chi^{2}=-3.540$ and $\Delta \rm AIC=4.460$, are considerably less supported by the CMB+BAO+SNe data combination. The current observations favor the IDE2+$N_{\rm eff}$ model with $\Delta \chi^{2}=-3.088$ and $\Delta \rm AIC=2.912$, and the IDE2+$\sum m_{\nu}$ model with $\Delta \chi^{2}=-2.790$ and $\Delta \rm AIC=3.210$. But the $H_{0}$ tension is still up to $3\sigma$ level or more in these cases. This research confirm that the IDE model with the constant $w$ and other some special parameters can relieve the $H_{0}$ tension to some extent. But compared to the $\Lambda$CDM model, when the number of extra parameters are three or more, the model will not consistent with current observational data.

\begin{table*}[!htp]
\centering
\renewcommand{\arraystretch}{1.5}
\scalebox{0.9}[0.9]{%
\begin{tabular}{|c|c c c|}
\hline
$S_{8}$&IDE1 &IDE1+$w$ &IDE2  \\
\hline

CMB+BAO+SNe
                 &$0 .824\pm0.010$
            &$0 .827\pm0.011$
            &$0 .827\pm0.011$
                     \\

CMB+BAO+SNe+DES
                                              &$0 .811\pm0.009$
                                              &$0 .812\pm0.009$
                                              &$0 .811\pm0.010$
                                              \\

\hline
\end{tabular}}
\caption{ Fitting results of the parameter $S_{8}$ in the IDE1, IDE1+$w$, and IDE2 models using the CMB+BAO+SNe data and the CMB+BAO+SNe+DES data. }
\label{tab:3}
\end{table*}

\begin{table*}[!htp]
\centering
\renewcommand{\arraystretch}{1.5}
\scalebox{0.9}[0.9]{%
\begin{tabular}{|c|c c c c|}
\hline
$S_{8}$&IDE2+$N_{\rm eff}$ &IDE2+$\sum m_{\nu}$ &IDE2+$N_{\rm eff}$+$\sum m_{\nu}$&IDE2+$N_{\rm eff}$+$m_{\nu,\rm sterile}^{\rm eff}$  \\
\hline

CMB+BAO+SNe
                 &$0 .825\pm0.011$
                                          &$0 .824\pm0.011$
                                          &$0 .822\pm0.011$
                                          &$0 .812^{+0.021}_{-0.014}$
                     \\

CMB+BAO+SNe+DES
                                              &$0 .809^{+0.010}_{-0.009}$
                                              &$0 .809\pm0.010$
                                              &$0 .806\pm0.010$
                                              &$0 .787\pm0.015$
                                              \\

\hline
\end{tabular}}
\caption{ Fitting results of $S_{8}$ in the IDE2+$N_{\rm eff}$, IDE2+$\sum m_{\nu}$, IDE2+$N_{\rm eff}$+$\sum m_{\nu}$, and IDE2+$N_{\rm eff}$+$m_{\nu,\rm sterile}^{\rm eff}$ models using the CMB+BAO+SNe data and the CMB+BAO+SNe+DES data.}
\label{tab:4}
\end{table*}

Instead of the matter fluctuation amplitude parameter $\sigma_{8}$, the related parameter $S_{8}\equiv\sigma_{8}\sqrt{\Omega_{\rm m}/0.3}$ is shown in Tab.\ref{tab:3} and Tab.\ref{tab:4}. Obviously, the central values for $S_{8}$ in the IDE models using the CMB+BAO+SNe data are larger than the DES Y1 best-fit values of $S_{8}\equiv 0.783^{+0.021}_{-0.025}$~\cite{DES:2017myr}, which means the existence of the $S_{8}$ (or $\sigma_{8}$) tension between the Planck data and the DES Y1 results. Furthermore, the DES Y1 data~\cite{DES:2017myr} including a total of 457 data points (abbreviated as the DES data) are combined with the CMB+BAO+SNe data. We find that the fitting values of $S_{8}$ from the CMB+BAO+SNe+DES data are lower than those derived from the CMB+BAO+SNe data, indicating that adding the DES data can decrease the fitting values of $S_{8}$ in these IDE models, thus alleviating the $S_{8}$ tension. In this paper, we only give a brief analysis on the fitting results of $S_{8}$ in these IDE models. The detailed analysis on this issue will be shown in a forthcoming longer paper.

\section{Conclusion}\label{sec:4}

We constrain the IDE1 model with $Q=\beta H \rho_{\rm c}$ and $w=-1$, the IDE1+$w$ model with $Q=\beta H \rho_{\rm c}$ and $w=constant$, and the IDE2 model with $Q=\beta wH \rho_{\rm c}$ and $w=constant$, by using CMB+BAO+SNe data combination. We find that the fitting value of $\beta$ is larger in the IDE2 model, but is smaller in the IDE1+$w$ model, than that in the IDE1 model. It indicates that the forms of $w$ and $Q$ can affect the fitting result of $\beta$ in IDE models. Comparing the fitting results of $H_{0}$ in the three IDE models, we quantitatively show the capability of these models to relieve the $H_{0}$ tension. We find that they all favor larger fitting values of $H_{0}$, thus reducing the discrepancy of $H_{0}$ between the Planck data and the direct $H_{0}$ measurement. Among them, the $H_{0}$ tension is about at the $3.5\sigma$ level for the IDE1+$w$ model and the IDE2 model, which favor larger fitting values of $H_{0}$ than that in the IDE1 model with $w=-1$. When the $\Lambda$CDM model is acted as a reference, the $H_{0}$ tensions are decreased by 7.03\%, 17.80\%, and 17.10\% in the three models. However, from a statistical point of view, adding a factor of the constant $w$ in the function of $Q=\beta H \rho_{\rm c}$ (i.e., the IDE2 model) is more consistent with current observational data than the case of the IDE1+$w$ model with $Q=\beta H \rho_{\rm c}$ and $w=constant$.

To further increase the fitting value of $H_{0}$ and reduce the discrepancy between the Planck data and the direct $H_{0}$ measurement, we investigate the constraint results of the IDE2+$N_{\rm eff}$ model, the IDE2+$\sum m_{\nu}$ model, the IDE2+$N_{\rm eff}$+$\sum m_{\nu}$ model, and the IDE2+$N_{\rm eff}$+$m_{\nu,\rm sterile}^{\rm eff}$ model. We find that adding these cosmological parameters in the IDE2 model actually increases the fitting value of $H_{0}$, and further reduces the $H_{0}$ tension. In particular, the $H_{0}$ tension can be decreased by 26.70\% in the IDE2+$N_{\rm eff}$+$m_{\nu,\rm sterile}^{\rm eff}$ model. However, when the values of $\chi^{2}$ and $\rm AIC$ in these models are discussed, we find that the model with more extra parameters is more inconsistent with the current observations. In the IDE2+$N_{\rm eff}$+$m_{\nu,\rm sterile}^{\rm eff}$ model, $\Delta \chi^{2}=-3.340$ and $\Delta \rm AIC=4.660$. Thus, none of these IDE models investigated in this paper can convincingly reconcile the tension of the Planck 2018 data with the direct measurement value of the Hubble constant.

\acknowledgments

This work was supported by the National Natural Science Foundation of China (Grant No. 12103038 and No. 11947022), the Natural Science Foundation of Shaanxi Provincial Department of Education (Grant No.20JK0683),  the Natural Science Foundation of Liaoning Province (Grant No. 2021-BS-154), and the 2019 Annual Scientific Research Funding Project of the Education Department of Liaoning Province (Grant No. LJC201915).


\begin{thebibliography}{99}

\bibitem{Aghanim:2018eyx}
  N.~Aghanim {\it et al.} [Planck Collaboration],
  ``Planck 2018 results. VI. cosmological parameters,"
  Astron.\ Astrophys.\  {\bf 641}, A6 (2020)
  doi = 10.1051/0004-6361/201833910
  [arXiv:1807.06209 [astro-ph.CO]].

\bibitem{Riess:2020fzl}
A.~G.~Riess, S.~Casertano, W.~Yuan, J.~B.~Bowers, L.~Macri, J.~C.~Zinn and D.~Scolnic,
``Cosmic distances calibrated to 1\% precision with Gaia EDR3 parallaxes and Hubble Space Telescope photometry of 75 milky way cepheids confirm tension with $\Lambda$CDM,''
Astrophys. J. Lett. \textbf{908}, no.1, L6 (2021)
doi:10.3847/2041-8213/abdbaf
[arXiv:2012.08534 [astro-ph.CO]].

\bibitem{Efstathiou:2013via}
G.~Efstathiou,
``$H_{0}$ revisited,''
Mon. Not. Roy. Astron. Soc. \textbf{440}, no.2, 1138-1152 (2014)
doi:10.1093/mnras/stu278
[arXiv:1311.3461 [astro-ph.CO]].

\bibitem{Spergel:2013rxa}
D.~N.~Spergel, R.~Flauger and R.~Hlo\v{z}ek,
``Planck data reconsidered,''
Phys. Rev. D \textbf{91}, no.2, 023518 (2015)
doi:10.1103/PhysRevD.91.023518
[arXiv:1312.3313 [astro-ph.CO]].

\bibitem{Addison:2015wyg}
G.~E.~Addison, Y.~Huang, D.~J.~Watts, C.~L.~Bennett, M.~Halpern, G.~Hinshaw and J.~L.~Weiland,
``Quantifying discordance in the 2015 Planck CMB spectrum,''
Astrophys. J. \textbf{818}, no.2, 132 (2016)
doi:10.3847/0004-637X/818/2/132
[arXiv:1511.00055 [astro-ph.CO]].

\bibitem{Planck:2016tof}
N.~Aghanim \textit{et al.} [Planck],
``Planck intermediate results. LI. features in the cosmic microwave background temperature power spectrum and shifts in cosmological parameters,''
Astron. Astrophys. \textbf{607}, A95 (2017)
doi:10.1051/0004-6361/201629504
[arXiv:1608.02487 [astro-ph.CO]].



\bibitem{Cardona:2016ems}
W.~Cardona, M.~Kunz and V.~Pettorino,
``Determining $H_0$ with bayesian hyper-parameters,''
JCAP \textbf{03}, 056 (2017)
doi:10.1088/1475-7516/2017/03/056
[arXiv:1611.06088 [astro-ph.CO]].

\bibitem{Zhang:2017aqn}
B.~R.~Zhang, M.~J.~Childress, T.~M.~Davis, N.~V.~Karpenka, C.~Lidman, B.~P.~Schmidt and M.~Smith,
``A blinded determination of $H_0$ from low-redshift Type Ia supernovae, calibrated by cepheid variables,''
Mon. Not. Roy. Astron. Soc. \textbf{471}, no.2, 2254-2285 (2017)
doi:10.1093/mnras/stx1600
[arXiv:1706.07573 [astro-ph.CO]].

\bibitem{Follin:2017ljs}
B.~Follin and L.~Knox,
``Insensitivity of the distance ladder Hubble constant determination to cepheid calibration modelling choices,''
Mon. Not. Roy. Astron. Soc. \textbf{477}, no.4, 4534-4542 (2018)
doi:10.1093/mnras/sty720
[arXiv:1707.01175 [astro-ph.CO]].

\bibitem{DiValentino:2021izs}
E.~Di Valentino, O.~Mena, S.~Pan, L.~Visinelli, W.~Yang, A.~Melchiorri, D.~F.~Mota, A.~G.~Riess and J.~Silk,
``In the realm of the Hubble tension\textemdash{}a review of solutions,''
Class. Quant. Grav. \textbf{38}, no.15, 153001 (2021)
doi:10.1088/1361-6382/ac086d
[arXiv:2103.01183 [astro-ph.CO]].

\bibitem{Freese:2004vs}
K.~Freese and D.~Spolyar,
``Chain inflation: bubble bubble toil and trouble,''
JCAP \textbf{07}, 007 (2005)
doi:10.1088/1475-7516/2005/07/007
[arXiv:hep-ph/0412145 [hep-ph]].



\bibitem{Li:2013dha}
M.~Li, X.~D.~Li, Y.~Z.~Ma, X.~Zhang and Z.~Zhang,
``Planck constraints on holographic dark energy,''
JCAP \textbf{09}, 021 (2013)
doi:10.1088/1475-7516/2013/09/021
[arXiv:1305.5302 [astro-ph.CO]].

\bibitem{Huang:2016fxc}
Q.~G.~Huang and K.~Wang,
``How the dark energy can reconcile Planck with local determination of the Hubble constant,''
Eur. Phys. J. C \textbf{76}, no.9, 506 (2016)
doi:10.1140/epjc/s10052-016-4352-x
[arXiv:1606.05965 [astro-ph.CO]].

\bibitem{Camarena:2018nbr}
D.~Camarena and V.~Marra,
``Impact of the cosmic variance on $H_0$ on cosmological analyses,''
Phys. Rev. D \textbf{98}, no.2, 023537 (2018)
doi:10.1103/PhysRevD.98.023537
[arXiv:1805.09900 [astro-ph.CO]].

\bibitem{Martinelli:2019krf}
M.~Martinelli and I.~Tutusaus,
``CMB tensions with low-redshift $H_0$ and $S_8$ measurements: impact of a redshift-dependent type-Ia supernovae intrinsic luminosity,''
Symmetry \textbf{11}, no.8, 986 (2019)
doi:10.3390/sym11080986
[arXiv:1906.09189 [astro-ph.CO]].

\bibitem{Planck:2018vyg}
N.~Aghanim \textit{et al.} [Planck],
``Planck 2018 results. VI. cosmological parameters,''
Astron. Astrophys. \textbf{641}, A6 (2020)
[erratum: Astron. Astrophys. \textbf{652}, C4 (2021)]
doi:10.1051/0004-6361/201833910
[arXiv:1807.06209 [astro-ph.CO]].

\bibitem{Yang:2021flj}
W.~Yang, E.~Di Valentino, S.~Pan, Y.~Wu and J.~Lu,
``Dynamical dark energy after Planck CMB final release and $H_0$ tension,''
Mon. Not. Roy. Astron. Soc. \textbf{501}, no.4, 5845-5858 (2021)
doi:10.1093/mnras/staa3914
[arXiv:2101.02168 [astro-ph.CO]].

\bibitem{Vagnozzi:2019ezj}
S.~Vagnozzi,
``New physics in light of the $H_0$ tension: an alternative view,''
Phys. Rev. D \textbf{102}, no.2, 023518 (2020)
doi:10.1103/PhysRevD.102.023518
[arXiv:1907.07569 [astro-ph.CO]].

\bibitem{Wang:2016lxa}
B.~Wang, E.~Abdalla, F.~Atrio-Barandela and D.~Pavon,
``Dark matter and dark energy interactions: theoretical challenges, cosmological implications and observational signatures,''
Rept. Prog. Phys. \textbf{79}, no.9, 096901 (2016)
doi:10.1088/0034-4885/79/9/096901
[arXiv:1603.08299 [astro-ph.CO]].

\bibitem{DiValentino:2017iww}
E.~Di Valentino, A.~Melchiorri and O.~Mena,
``Can interacting dark energy solve the $H_0$ tension?,''
Phys. Rev. D \textbf{96}, no.4, 043503 (2017)
doi:10.1103/PhysRevD.96.043503
[arXiv:1704.08342 [astro-ph.CO]].

\bibitem{Yang:2017ccc}
W.~Yang, S.~Pan and D.~F.~Mota,
``Novel approach toward the large-scale stable interacting dark-energy models and their astronomical bounds,''
Phys. Rev. D \textbf{96}, no.12, 123508 (2017)
doi:10.1103/PhysRevD.96.123508
[arXiv:1709.00006 [astro-ph.CO]].


\bibitem{Yang:2017zjs}
W.~Yang, S.~Pan and J.~D.~Barrow,
``Large-scale stability and astronomical constraints for coupled dark-energy models,''
Phys. Rev. D \textbf{97}, no.4, 043529 (2018)
doi:10.1103/PhysRevD.97.043529
[arXiv:1706.04953 [astro-ph.CO]].


\bibitem{Yang:2018euj}
W.~Yang, S.~Pan, E.~Di Valentino, R.~C.~Nunes, S.~Vagnozzi and D.~F.~Mota,
``Tale of stable interacting dark energy, observational signatures, and the $H_0$ tension,''
JCAP \textbf{09}, 019 (2018)
doi:10.1088/1475-7516/2018/09/019
[arXiv:1805.08252 [astro-ph.CO]].

\bibitem{Yang:2018xlt}
W.~Yang, S.~Pan, R.~Herrera and S.~Chakraborty,
``Large-scale (in) stability analysis of an exactly solved coupled dark-energy model,''
Phys. Rev. D \textbf{98}, no.4, 043517 (2018)
doi:10.1103/PhysRevD.98.043517
[arXiv:1808.01669 [gr-qc]].

\bibitem{Feng:2017usu}
L.~Feng, J.~F.~Zhang and X.~Zhang,
``Search for sterile neutrinos in a universe of vacuum energy interacting with cold dark matter,''
Phys. Dark Univ. \textbf{23}, 100261 (2019)
doi:10.1016/j.dark.2018.100261
[arXiv:1712.03148 [astro-ph.CO]].

\bibitem{An:2018vzw}
R.~An, A.~A.~Costa, L.~Xiao, J.~Zhang and B.~Wang,
``Testing a quintessence model with Yukawa interaction from cosmological observations and N-body simulations,''
Mon. Not. Roy. Astron. Soc. \textbf{489}, no.1, 297-309 (2019)
doi:10.1093/mnras/stz2028
[arXiv:1809.03224 [astro-ph.CO]].

\bibitem{Li:2019ajo}
H.~L.~Li, D.~Z.~He, J.~F.~Zhang and X.~Zhang,
``Quantifying the impacts of future gravitational-wave data on constraining interacting dark energy,''
JCAP \textbf{06}, 038 (2020)
doi:10.1088/1475-7516/2020/06/038
[arXiv:1908.03098 [astro-ph.CO]].

\bibitem{DiValentino:2019ffd}
E.~Di Valentino, A.~Melchiorri, O.~Mena and S.~Vagnozzi,
``Interacting dark energy in the early 2020s: a promising solution to the $H_0$ and cosmic shear tensions,''
Phys. Dark Univ. \textbf{30}, 100666 (2020)
doi:10.1016/j.dark.2020.100666
[arXiv:1908.04281 [astro-ph.CO]].

\bibitem{DiValentino:2019jae}
E.~Di Valentino, A.~Melchiorri, O.~Mena and S.~Vagnozzi,
``Nonminimal dark sector physics and cosmological tensions,''
Phys. Rev. D \textbf{101}, no.6, 063502 (2020)
doi:10.1103/PhysRevD.101.063502
[arXiv:1910.09853 [astro-ph.CO]].

\bibitem{Yang:2021oxc}
W.~Yang, S.~Pan, L.~Arest\'e Sal\'o and J.~de Haro,
``Theoretical and observational bounds on some interacting vacuum energy scenarios,''
Phys. Rev. D \textbf{103}, no.8, 083520 (2021)
doi:10.1103/PhysRevD.103.083520
[arXiv:2104.04505 [astro-ph.CO]].

\bibitem{Yang:2021hxg}
W.~Yang, S.~Pan, E.~Di Valentino, O.~Mena and A.~Melchiorri,
``2021-$H_0$ odyssey: closed, phantom and interacting dark energy cosmologies,''
[arXiv:2101.03129 [astro-ph.CO]].







\bibitem{DiValentino:2020kpf}
E.~Di Valentino, A.~Melchiorri, O.~Mena, S.~Pan and W.~Yang,
``Interacting dark energy in a closed universe,''
Mon. Not. Roy. Astron. Soc. \textbf{502}, no.1, L23-L28 (2021)
doi:10.1093/mnrasl/slaa207
[arXiv:2011.00283 [astro-ph.CO]].

\bibitem{Zhang:2021yof}
M.~Zhang, B.~Wang, P.~J.~Wu, J.~Z.~Qi, Y.~Xu, J.~F.~Zhang and X.~Zhang,
``Prospects for constraining interacting dark energy models with 21 cm intensity mapping experiments,''
Astrophys. J. \textbf{918}, no.2, 56 (2021)
doi:10.3847/1538-4357/ac0ef5
[arXiv:2102.03979 [astro-ph.CO]].

\bibitem{Bonilla:2021dql}
A.~Bonilla, S.~Kumar, R.~C.~Nunes and S.~Pan,
``Reconstruction of the dark sectors' interaction: a model-independent inference and forecast from GW standard sirens,''
[arXiv:2102.06149 [astro-ph.CO]].

\bibitem{Johnson:2021wou}
J.~P.~Johnson, A.~Sangwan and S.~Shankaranarayanan,
``Cosmological perturbations in the interacting dark sector: observational constraints and predictions,''
[arXiv:2102.12367 [astro-ph.CO]].

\bibitem{Battye:2013xqa}
R.~A.~Battye and A.~Moss,
``Evidence for massive neutrinos from cosmic microwave background and lensing observations,''
Phys. Rev. Lett. \textbf{112}, no.5, 051303 (2014)
doi:10.1103/PhysRevLett.112.051303
[arXiv:1308.5870 [astro-ph.CO]].

\bibitem{SDSS:2014iwm}
M.~Betoule \textit{et al.} [SDSS],
``Improved cosmological constraints from a joint analysis of the SDSS-II and SNLS supernova samples,''
Astron. Astrophys. \textbf{568}, A22 (2014)
doi:10.1051/0004-6361/201423413
[arXiv:1401.4064 [astro-ph.CO]].


\bibitem{Bernal:2016gxb}
J.~L.~Bernal, L.~Verde and A.~G.~Riess,
``The trouble with $H_0$,''
JCAP \textbf{10}, 019 (2016)
doi:10.1088/1475-7516/2016/10/019
[arXiv:1607.05617 [astro-ph.CO]].

\bibitem{Guo:2017hea}
R.~Y.~Guo, Y.~H.~Li, J.~F.~Zhang and X.~Zhang,
``Weighing neutrinos in the scenario of vacuum energy interacting with cold dark matter: application of the parameterized post-Friedmann approach,''
JCAP \textbf{05}, 040 (2017)
doi:10.1088/1475-7516/2017/05/040
[arXiv:1702.04189 [astro-ph.CO]].

\bibitem{Guo:2017qjt}
R.~Y.~Guo and X.~Zhang,
``Constraints on inflation revisited: an analysis including the latest local measurement of the Hubble constant,''
Eur. Phys. J. C \textbf{77}, no.12, 882 (2017)
doi:10.1140/epjc/s10052-017-5454-9
[arXiv:1704.04784 [astro-ph.CO]].

\bibitem{Zhao:2017urm}
M.~M.~Zhao, D.~Z.~He, J.~F.~Zhang and X.~Zhang,
``Search for sterile neutrinos in holographic dark energy cosmology: reconciling Planck observation with the local measurement of the Hubble constant,''
Phys. Rev. D \textbf{96}, no.4, 043520 (2017)
doi:10.1103/PhysRevD.96.043520
[arXiv:1703.08456 [astro-ph.CO]].

\bibitem{Guo:2018gyo}
R.~Y.~Guo, J.~F.~Zhang and X.~Zhang,
``Exploring neutrino mass and mass hierarchy in the scenario of vacuum energy interacting with cold dark matte,''
Chin. Phys. C \textbf{42}, no.9, 095103 (2018)
doi:10.1088/1674-1137/42/9/095103
[arXiv:1803.06910 [astro-ph.CO]].

\bibitem{DiValentino:2017rcr}
E.~Di Valentino, E.~V.~Linder and A.~Melchiorri,
``Vacuum phase transition solves the $H_0$ tension,''
Phys. Rev. D \textbf{97}, no.4, 043528 (2018)
doi:10.1103/PhysRevD.97.043528
[arXiv:1710.02153 [astro-ph.CO]].

\bibitem{Guo:2018ans}
R.~Y.~Guo, J.~F.~Zhang and X.~Zhang,
``Can the $H_0$ tension be resolved in extensions to $\Lambda$CDM cosmology?,''
JCAP \textbf{02}, 054 (2019)
doi:10.1088/1475-7516/2019/02/054
[arXiv:1809.02340 [astro-ph.CO]].


\bibitem{Yang:2020tax}
W.~Yang, E.~Di Valentino, O.~Mena and S.~Pan,
``Dynamical dark sectors and neutrino masses and abundances,''
Phys. Rev. D \textbf{102}, no.2, 023535 (2020)
doi:10.1103/PhysRevD.102.023535
[arXiv:2003.12552 [astro-ph.CO]].



\bibitem{Yang:2020ope}
W.~Yang, E.~Di Valentino, S.~Pan and O.~Mena,
``Emergent dark energy, neutrinos and cosmological tensions,''
Phys. Dark Univ. \textbf{31}, 100762 (2021)
doi:10.1016/j.dark.2020.100762
[arXiv:2007.02927 [astro-ph.CO]].

\bibitem{Feng:2019jqa}
L.~Feng, D.~Z.~He, H.~L.~Li, J.~F.~Zhang and X.~Zhang,
``Constraints on active and sterile neutrinos in an interacting dark energy cosmology,''
Sci. China Phys. Mech. Astron. \textbf{63}, no.9, 290404 (2020)
doi:10.1007/s11433-019-1511-8
[arXiv:1910.03872 [astro-ph.CO]].

\bibitem{Poulin:2018cxd}
V.~Poulin, T.~L.~Smith, T.~Karwal and M.~Kamionkowski,
``Early dark energy can resolve the Hubble tension,''
Phys. Rev. Lett. \textbf{122}, no.22, 221301 (2019)
doi:10.1103/PhysRevLett.122.221301
[arXiv:1811.04083 [astro-ph.CO]].

\bibitem{Sakstein:2019fmf}
J.~Sakstein and M.~Trodden,
``Early dark energy from massive neutrinos as a natural resolution of the Hubble tension,''
Phys. Rev. Lett. \textbf{124}, no.16, 161301 (2020)
doi:10.1103/PhysRevLett.124.161301
[arXiv:1911.11760 [astro-ph.CO]].

\bibitem{Li:2014cee}
  Y.~H.~Li, J.~F.~Zhang and X.~Zhang,
  ``Exploring the full parameter space for an interacting dark energy model with recent observations including redshift-space distortions: application of the Parametrized Post-Friedmann approach,"
  Phys.\ Rev.\ D {\bf 90}, no. 12, 123007 (2014)
  doi:10.1103/PhysRevD.90.123007
  [arXiv:1409.7205 [astro-ph.CO]].

\bibitem{Li:2015vla}
Y.~H.~Li, J.~F.~Zhang and X.~Zhang,
``Testing models of vacuum energy interacting with cold dark matter,''
Phys. Rev. D \textbf{93}, no.2, 023002 (2016)
doi:10.1103/PhysRevD.93.023002
[arXiv:1506.06349 [astro-ph.CO]].

\bibitem{Anchordoqui:2021gji}
L.~A.~Anchordoqui, E.~Di Valentino, S.~Pan and W.~Yang,
``Dissecting the $H_{0}$ and $S_{8}$ tensions with Planck + BAO + supernova type Ia in multi-parameter cosmologies,''
JHEAp \textbf{32}, 28-64 (2021)
doi:10.1016/j.jheap.2021.08.001
[arXiv:2107.13932 [astro-ph.CO]].



\bibitem{Pan:2019gop}
S.~Pan, W.~Yang, E.~Di Valentino, E.~N.~Saridakis and S.~Chakraborty,
``Interacting scenarios with dynamical dark energy: observational constraints and alleviation of the $H_0$ tension,''
Phys. Rev. D \textbf{100}, no.10, 103520 (2019)
doi:10.1103/PhysRevD.100.103520
[arXiv:1907.07540 [astro-ph.CO]].

\bibitem{Valiviita:2008iv}
J.~Valiviita, E.~Majerotto and R.~Maartens,
``Instability in interacting dark energy and dark matter fluids,''
JCAP \textbf{07}, 020 (2008)
doi:10.1088/1475-7516/2008/07/020
[arXiv:0804.0232 [astro-ph]].

\bibitem{Benetti:2019lxu}
M.~Benetti, W.~Miranda, H.~A.~Borges, C.~Pigozzo, S.~Carneiro and J.~S.~Alcaniz,
``Looking for interactions in the cosmological dark sector,''
JCAP \textbf{12}, 023 (2019)
doi:10.1088/1475-7516/2019/12/023
[arXiv:1908.07213 [astro-ph.CO]].

\bibitem{Yang:2019uog}
W.~Yang, S.~Pan, R.~C.~Nunes and D.~F.~Mota,
``Dark calling dark: interaction in the dark sector in presence of neutrino properties after Planck CMB final release,''
JCAP \textbf{04}, 008 (2020)
doi:10.1088/1475-7516/2020/04/008
[arXiv:1910.08821 [astro-ph.CO]].




\bibitem{Bento:2002yx}
M.~d.~Bento, O.~Bertolami and A.~A.~Sen,
``Generalized chaplygin gas and CMBR constraints,''
Phys. Rev. D \textbf{67}, 063003 (2003)
doi:10.1103/PhysRevD.67.063003
[arXiv:astro-ph/0210468 [astro-ph]].

\bibitem{Zhang:2004gc}
X.~Zhang, F.~Q.~Wu and J.~Zhang,
``A new generalized chaplygin gas as a scheme for unification of dark energy and dark matter,''
JCAP \textbf{01}, 003 (2006)
doi:10.1088/1475-7516/2006/01/003
[arXiv:astro-ph/0411221 [astro-ph]].

\bibitem{Wang:2013qy}
Y.~Wang, D.~Wands, L.~Xu, J.~De-Santiago and A.~Hojjati,
``Cosmological constraints on a decomposed chaplygin gas,''
Phys. Rev. D \textbf{87}, no.8, 083503 (2013)
doi:10.1103/PhysRevD.87.083503
[arXiv:1301.5315 [astro-ph.CO]].




\bibitem{Hu:2008zd}
  W.~Hu,
  ``Parametrized post-friedmann signatures of acceleration in the CMB,"
  Phys.\ Rev.\ D {\bf 77}, 103524 (2008)
  doi:10.1103/PhysRevD.77.103524
  [arXiv:0801.2433 [astro-ph]].


\bibitem{Fang:2008sn}
  W.~Fang, W.~Hu and A.~Lewis,
  ``Crossing the phantom divide with parameterized post-friedmann dark energy,"
  Phys.\ Rev.\ D {\bf 78}, 087303 (2008)
  doi:10.1103/PhysRevD.78.087303
  [arXiv:0808.3125 [astro-ph]].

\bibitem{Majerotto:2009zz}
E.~Majerotto, J.~Valiviita and R.~Maartens,
``Instability in interacting dark energy and dark matter fluids,''
Nucl. Phys. B Proc. Suppl. \textbf{194}, 260-265 (2009)
doi:10.1016/j.nuclphysbps.2009.07.089

\bibitem{He:2008si}
J.~H.~He, B.~Wang and E.~Abdalla,
``Stability of the curvature perturbation in dark sectors' mutual interacting models,''
Phys. Lett. B \textbf{671}, 139-145 (2009)
doi:10.1016/j.physletb.2008.11.062
[arXiv:0807.3471 [gr-qc]].

\bibitem{Clemson:2011an}
T.~Clemson, K.~Koyama, G.~B.~Zhao, R.~Maartens and J.~Valiviita,
``Interacting dark energy -- constraints and degeneracies,''
Phys. Rev. D \textbf{85}, 043007 (2012)
doi:10.1103/PhysRevD.85.043007
[arXiv:1109.6234 [astro-ph.CO]].

\bibitem{Li:2014eha}
  Y.~H.~Li, J.~F.~Zhang and X.~Zhang,
  ``Parametrized post-friedmann framework for interacting dark energy,"
  Phys.\ Rev.\ D {\bf 90}, no. 6, 063005 (2014)
  doi:10.1103/PhysRevD.90.063005
  [arXiv:1404.5220 [astro-ph.CO]].




\bibitem{Beutler:2011hx}
  F.~Beutler {\it et al.},
  ``The 6dF galaxy survey: baryon acoustic oscillations and the local hubble constant,"
  Mon.\ Not.\ Roy.\ Astron.\ Soc.\  {\bf 416}, 3017 (2011).
  [arXiv:1106.3366 [astro-ph.CO]].


\bibitem{Ross:2014qpa}
  A.~J.~Ross, L.~Samushia, C.~Howlett, W.~J.~Percival, A.~Burden and M.~Manera,
  ``The clustering of the SDSS DR7 main galaxy sample I: a 4 percent distance measure at $z = 0.15$,''
  Mon.\ Not.\ Roy.\ Astron.\ Soc.\  {\bf 449}, no. 1, 835 (2015).
  doi:10.1093/mnras/stv154
  [arXiv:1409.3242 [astro-ph.CO]].


\bibitem{Alam:2016hwk}
  S.~Alam {\it et al.} [BOSS Collaboration],
  ``The clustering of galaxies in the completed SDSS-III baryon oscillation spectroscopic survey: cosmological analysis of the DR12 galaxy sample,"
  Mon.\ Not.\ Roy.\ Astron.\ Soc.\  {\bf 470}, no. 3, 2617 (2017)
  doi:10.1093/mnras/stx721
  [arXiv:1607.03155 [astro-ph.CO]].

\bibitem{Scolnic:2017caz}
  D.~M.~Scolnic {\it et al.},
  ``The complete light-curve sample of spectroscopically confirmed SNe Ia from Pan-STARRS1 and cosmological constraints from the combined pantheon sample,"
  Astrophys.\ J.\  {\bf 859}, no. 2, 101 (2018)
  doi:10.3847/1538-4357/aab9bb
  [arXiv:1710.00845 [astro-ph.CO]].

\bibitem{H.Akaike:1974}
 H. Akaike, ``A new look at the statistical model identication," IEEE Trans. Automatic Control 19, 716 (1974).

\bibitem{Lewis:1999bs}
A.~Lewis, A.~Challinor and A.~Lasenby,
``Efficient computation of CMB anisotropies in closed FRW models,''
Astrophys. J. \textbf{538}, 473-476 (2000)
doi:10.1086/309179
[arXiv:astro-ph/9911177 [astro-ph]].

\bibitem{Lewis:2013hha}
A.~Lewis,
``Efficient sampling of fast and slow cosmological parameters,''
Phys. Rev. D \textbf{87}, no.10, 103529 (2013)
doi:10.1103/PhysRevD.87.103529
[arXiv:1304.4473 [astro-ph.CO]].



\bibitem{Riess:2019cxk}
A.~G.~Riess, S.~Casertano, W.~Yuan, L.~M.~Macri and D.~Scolnic,
``Large magellanic cloud cepheid standards provide a 1\% foundation for the determination of the Hubble constant and stronger evidence for physics beyond $\Lambda$CDM,''
Astrophys. J. \textbf{876}, no.1, 85 (2019)
doi:10.3847/1538-4357/ab1422
[arXiv:1903.07603 [astro-ph.CO]].

\bibitem{DES:2017myr}
T.~M.~C.~Abbott \textit{et al.} [DES],
``Dark energy survey year 1 results: cosmological constraints from galaxy clustering and weak lensing,''
Phys. Rev. D \textbf{98}, no.4, 043526 (2018)
doi:10.1103/PhysRevD.98.043526
[arXiv:1708.01530 [astro-ph.CO]].



\end{thebibliography}
\end{document}